\documentclass{article}
\usepackage{float,fullpage,graphicx}
\usepackage{amsmath}
\usepackage{hyperref}
\usepackage{pifont}
\newcommand{\cmark}{\ding{51}}%
\newcommand{\xmark}{\ding{55}}
\title{Fast Search with Poor OCR}
\author{Taivanbat Badamdorj\footnote{Research conducted while at Tel Aviv University.} \\
University of Alberta \\ 
\and Adiel Ben-Shalom \\ 
Tel Aviv University \\ 
\and Nachum Dershowitz \\
Tel Aviv University \\ 
\and Lior Wolf \\ 
Facebook AI Research 
and Tel Aviv University \\ 
}

\begin{document}
\maketitle

\begin{abstract}
The indexing and searching of historical documents have garnered attention in recent years due to massive digitization efforts of important collections worldwide. Pure textual search in these corpora is a problem since optical character recognition (OCR) is infamous for performing poorly on such historical material, which often suffer from poor preservation. We propose a novel text-based method for searching through noisy text. 
Our system  represents words as vectors, projects queries and candidates obtained from the OCR into a common space, and ranks the candidates using a metric suited to nearest-neighbor search. We demonstrate the practicality of our method on typewritten German documents from the WWII era. 

\end{abstract}

\section{Introduction}

The Wiener Library is one of the most extensive archives on the Holocaust and Nazi era. 
Established in the 1930s, the library's unique collection of over one million items includes press cuttings, eyewitness testimonies, photographs, as well as published and unpublished works from that era. 

These documents have been  digitized and made available online. Hence, it's of great importance to have a fast search tool for them.

We present an easy-to-implement method that enables scholars and the public to effectively search such large collections of textual material that would otherwise be inaccessible due to the lack of reliable transcriptions. 
Its main advantage is its simplicity, in contrast to existing image-based methods  (such as various word-spotting methods \cite{almazan,manmatha,sudholt,yalniz}), which are technically involved and need to be retrained and/or rebuilt for each script style within a dataset. 
Our system is based on existing optical character recognition (OCR) tools that work well for a variety of typefaces and languages. The ease and low cost of implementation enables anyone to apply the method to various datasets without substantial expertise in computer vision.

Image-based search systems \cite{almazan,manmatha,sudholt,yalniz} often encode the query string and candidate images into a common subspace, and find matches using nearest neighbors. 
Analogously, we encode the noisy candidate outputs provided by the OCR engine as well as the query in a fixed vector representation, and learn a common space between them. 

At the same time, the size of the collection we are dealing with (over 70,000 documents) also makes efficiency a primary concern. Pure text-based distance metrics such as edit-distance are impractical for the task of searching through large and noisy corpora; our corpus has close to 2 million unique reads obtained by OCR.

Thus searching using nearest neighbors is an easy and quite common solution. This step is fast, since it can use efficient matrix multiplication implementations on modern GPUs \cite{fastnn}. Although the most common metric for nearest neighbors is simply their cosine distance, we use a new metric introduced in \cite{csls} that is still fast, and also improves retrieval results.

\section{The Corpus}
\subsection{History}
The Wiener Collection was first established in 1933 by Dr.\@ Alfred Wiener, a German Jewish scholar and former member and activist of the Centralverein deutscher Staatsb{\"u}rger j{\"u}dischen Glaubens. 
Wiener left Germany when the Nazis rose to power and established the Jewish Central Information Office (JCIO) in Amsterdam.
The idea was to collect information about the Nazi Party as part of the struggle to prevent its strengthening and to draw world attention towards the dangers of Nazi antisemitism.
In 1939, Wiener transferred the collection to London. 
Throughout the war years, Wiener and his assistants continued to collect information and documents on Germany's occupation policy, responses to it, and particularly on the fate of European Jewry. 
When the war ended, Holocaust survivors' testimonies, as well as information regarding the fate of Jewish refugees, were added. 
The  collection  played an important role in the charges leveled against war criminals and has continued to serve the media and  scholars.

\subsection{Holdings} 

The collection comprises publications on the Third Reich, during and between the two world wars, Jewish communities in Europe, the Holocaust, antisemitism, and fascism throughout the world.
These include approximately 150,000 books, reference works, pamphlets and journals; over one million indexed newspaper clippings, unpublished memoirs, and interviews; around 40,000 documents on the Nuremberg trials; various editions and extensive literature on \textit{The Protocols of the Elders of Zion}; dossiers on war criminals; documents on the ``Jewish Question'' taken from records of the Gestapo, the Reichskanzlei, and the Foreign Office of the Third Reich; more than 500 microfilm and microfiche titles; and over 300 subscriptions to journals, both Holocaust and extreme right wing/Holocaust denial.

\subsection{Online Archive} 

In the late 1970s, the Wiener Library (London) transferred most of its collection to Tel Aviv University in Israel,
where the Wiener Library for the Study of the Nazi Era and the Holocaust was established.
In addition, microfilms of parts of the collection are held  in London and Tel Aviv.

In recent years, the documents at both the London and Tel Aviv locations are being digitized and made available online.
Beginning in 2015, Tel Aviv digitized its ``500 Document Collection,'' which includes the main component of the original Wiener materials. 
It comprises  more than 75,000 images that were scanned from microfilm and microfiche reproductions of the originals. 
The work described in this paper was performed with this online collection. 

\section{Related Work}

Many recent projects have used word-spotting techniques to search  large textual corpora. 
Spotting was first introduced in \cite{manmatha} to enable search through handwritten documents, and was further improved upon in \cite{almazan,frinken,ws1,ws2,yalniz}. 
The work of \cite{almazan} is interesting because of the vector encoding for words that they introduced. 

More recently, \cite{krishnan} and \cite{sudholt}  introduced deep learning-based approaches to the task.
The former (\cite{krishnan}) jointly learned the text and image embeddings using a neural network. 
The latter (\cite{sudholt}) implicitly trained an OCR by training a convolutional neural network (CNN) to output its ground truth vector representation. 

Other works that incorporate searching through noisy OCR texts include \cite{collins-t,fataicha}, both of which use string-based methods to find possible matches. 
We found using purely string-based methods for matching to be too slow for our needs.

\section{Method}

Given a text query, we would like to find the correct matches among the noisy candidate output by the OCR. 
Our method consists of the following steps:
\begin{itemize}
    \item[(A)] preprocessing the images; \item[(B)] obtaining noisy candidate words using an OCR engine; \item[(C)] encoding the query and candidate words into vectors; \item[(D)] learning a common subspace between them; and finally \item[(E)] ranking the candidates according to distance from the query.
\end{itemize}

\subsection{Preprocessing} 

Information lost at the beginning of this pipeline is irrecoverable, so we take care to preprocess the images properly so that  OCR may perform optimally. 

The documents were photographed with a black border around each page. 
We first remove these borders  by binarizing the image and finding the largest connected component, which will be the black border. 
We find its boundaries and keep only that part that is enclosed inside, which is the actual document of interest. 

There is a stark difference in lighting between many of the documents. 
Some documents are very dark, such as in Figure~\ref{fig:dark_document}. 
Other documents are not uniformly lit and have light or dark patches. 
So we adjust the contrast using the CLAHE \cite{clahe} algorithm. 
CLAHE  adjusts the contrast in local regions, thus alleviating the problem of nonuniform lighting. 

Other possible steps, such as binarization \cite{otsu}, degraded the performance of the OCR across all tasks. 

\subsection{OCR} 

Encouraged by recent improvements in optical character recognition software, we use OCR to obtain candidates for search.
Of the many  options, we chose Tesseract \cite{smith, smith2},%
\footnote{Tesseract instructions and pretrained models are available at \url{https://github.com/tesseract-ocr}.} an open-source OCR engine,
for its quality and convenience. 
For each word, we use Tesseract to get its bounding box and its transcription. 

Tesseract also works well with a variety of fonts and languages \cite{smith3} as needed for this project. 
The latest version -- which is the one we used -- is based on long short-term memory \cite{hoch} with connectionist temporal classification (CTC) \cite{graves} used as the scoring function, a common combination in unaligned sequence generation problems.
We also used pytesseract,\footnote{pytesseract is available at \url{https://pypi.org/project/pytesseract}.} a Python wrapper that allowed us to interface with the Tesseract engine more easily.

\begin{figure*}[t]
  \centering
  \includegraphics[width=0.85\linewidth]{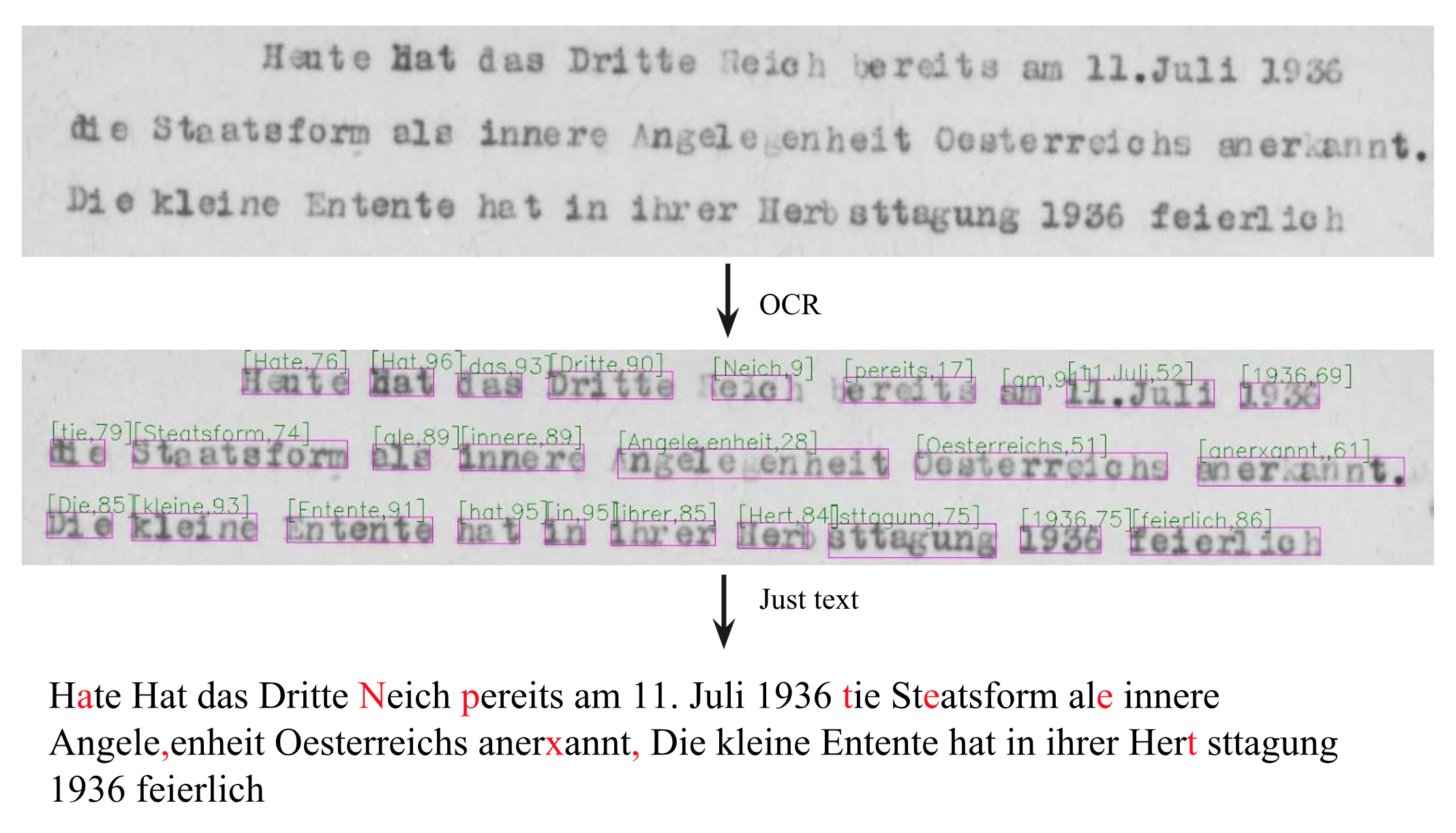}
  \caption{Example of Tesseract OCR output from best performing German model. Misreadings shown in red.} 
  \label{fig:after_ocr}
\end{figure*}





\subsection{Encoding} 

After obtaining the readings, we would like to have a fixed-size vector representation of each word to do quick nearest-neighbor search. We want the correct candidate vectors to be close to the query vector. 

To achieve this, we propose the use of the following vector representation. 

\subsubsection{PHOC} 

The pyramidal histogram of character (PHOC) encoding was introduced in \cite{almazan}, in the context of word-spotting. 
In the simplest case, words can be represented simply by the characters that are in the word, that is, by a vector where 1 indicates that a character is in the word, and 0 otherwise. 
In this case, however, the words ``beard'' and ``bread'' would have the exact same representation. 
Since the letter ``r'' is in the second half of the word ``beard'' and in the first half of the word ``bread'', we could use that to differentiate between the two words. 
Thus we define two more binary vectors, one for all the characters in the first half of the word, and another for the words in the second half of the word. 
Proceeding along the same lines, we could divide the word into thirds, quarters, and so on. 
We refer to the number of divisions as ``levels''. 
The final representation of the word is the concatenation of all its vector representations at each level. 

\begin{figure}[t]
  \centering
  \includegraphics[width=0.7\linewidth]{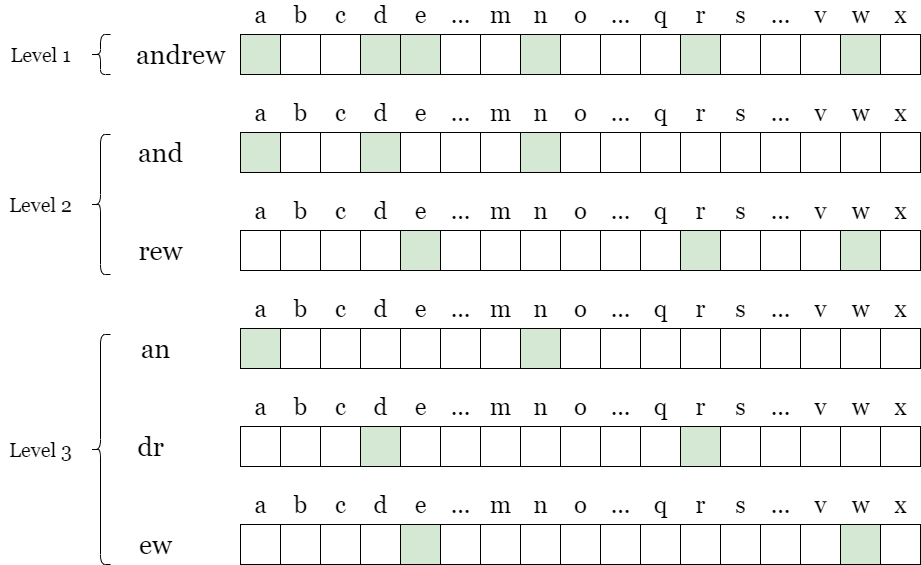}
  \caption{Illustration of PHOC. The final representation is the concatenation of all the binary vectors obtained at each level.} 
  \label{fig:PHOC}
\end{figure}

Formally, we decide where to assign each character in the following manner: 

We first define the normalized occupancy of the $k$th character $w_k$ of a word $w$ with length $n$ as $$Occ(k,n) = \left[\frac{k}{n}, \frac{k+1}{n}\right]$$ where the positions $k$ start from $0$. 
Using the same formula for the occupancy of region $r$ at level $\ell$, the character $w_k$ belongs to  region $r$ if the overlap between their respective occupancies is greater than or equal to $50\%$ of the occupancy area of the character, that is, if $$\frac{|Occ(k,n) \cap Occ(r,\ell)|}{|Occ(k,n)|} \geq \frac12$$ where $|[a,b]| = b - a$.

In simpler terms, were all characters  of  uniform width, in other words,  each occupying  unit area, and were we to divide an $n$-letter  word into $\ell$ bins,   each bin occupying an area of ${n}/{\ell}$ units, then we would assign a letter into each bin where at least half of the letter overlaps with the bin. 

\subsubsection{Character Set and Special Considerations}

Our dataset has additional special characters from multiple languages. 
Although most of our text is in German, there are also Polish, English, and even some Hebrew texts. 
We focus our efforts on the Latin-based languages, for which we have a set of 96 unique characters. 

The German language is also famous for its compound words, the practice of combining multiple words into one word, resulting in many long words in our corpus. 
Therefore, we use levels 1, 2, 4, and 8 for computing the PHOC histograms. 
The end result is that each word is represented by a binary vector of size 1440.

The vector representations of each word are precomputed and stored in memory. 
For our dataset, this amounts to a large 80GB matrix. 
But computing the distances for a single query with over 1 million candidates takes only 7 seconds using a GPU.

\begin{figure*}[t]
  \centering
  \includegraphics[width=1\linewidth]{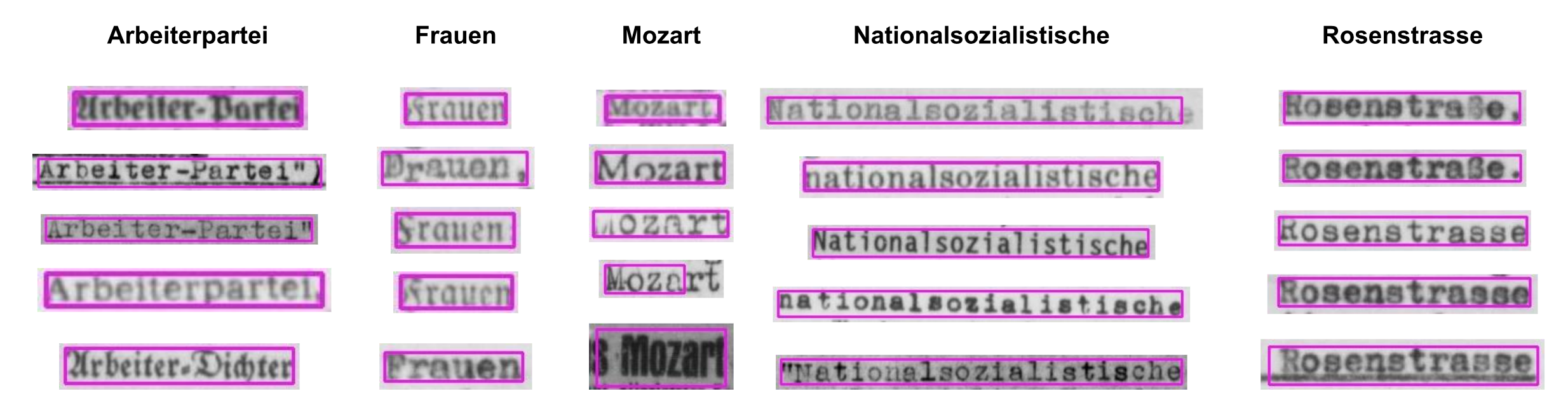}
  \caption{Qualitative search results for the entire dataset with bounding boxes found by the OCR engine in pink. Our model works well even for cases where there is substantial fading of the ink, and cases where the transcription of the candidate itself is not an exact match with the query.} 
  \label{fig:qualitative}
\end{figure*}

\subsection{Learning a Common Space using Canonical Correlation Analysis}

Canonical correlation analysis (CCA) \cite{hotelling} is a statistical method for computing a linear projection for two views into a common space by maximizing their correlation. It has been widely used in the field of computer vision to tackle tasks such as linking text to images, multiview analysis \cite{cca1}, and action recognition \cite{cca2}. 

Multiple variants have been proposed over the years: it was regularized \cite{vinod}, kernels were introduced \cite{kernel1, kernel2}, and more recently, deep learning methods were created \cite{deepcca1, deepcca2}. 

CCA takes two vectors, often referred to as ``views'' in the literature, $X_i$ and $Y_i$, as inputs. They are stacked as columns of two matrices $X$ and $Y$. 
The columns of matrices $X$ and $Y$ are assumed to be centered, that is, $\sum{X_i} = 0$ and $\sum{Y_i} = 0$. 
CCA learns a common space by learning two matrices $W_x$ and $W_y$ such that the correlation $\sum_i {(W_x X_i)^T (W_y Y_i)}$ is maximized under the condition  that the projection is uncorrelated, viz.

\begin{align*}
\sum_i {(W_x X_i)^T (W_x X_i)} &= I \\
\sum_i {(W_y Y_i)^T (W_y Y_i)} &= I
\end{align*}

We use the regularized version of CCA \cite{vinod}, which adds a regularization term to the covariance matrices. The regularized version generalizes better and is more stable. 

We consider the PHOC vector of the correct spelling of a word (the query) to be one view and the PHOC vector of its corrupted versions (the candidates) as another view. For each training fold, we use half of the pages to obtain the noisy candidates with their ground-truth transcriptions, computed their PHOC representations, and learned the matrices $W_x$ and $W_y$. Because the data is assumed to be centered, we also compute the mean of our candidates and ground-truth transcriptions during training, and subtract those means at query time.

\subsection{Ranking Candidates}

Having projected queries and candidates into a common subspace, we find the correct candidates by means of nearest-neighbor search. 

\subsubsection{Nearest Neighbors}

Finding the correct candidates for a given query by using nearest neighbors \cite{neighbor,neighbor2} is common in related tasks such as topic detection \cite{td}.

However, nearest-neighbor search is by its very nature asymmetric: $A$ being a nearest neighbor of $B$ does not imply that $B$ is also the nearest neighbor of $A$. 
In high-dimensional spaces, this leads to an effect that causes faulty matching when using a nearest-neighbor rule \cite{radovanovic}: Some vectors, dubbed ``hubs'', have a high probability of being the nearest neighbor of many other vectors, while other vectors, ``anti-hubs'', are not the nearest neighbors of any other points. 
Recently, methods such as inverted softmax (ISF) \cite{isf}, and cross-domain similarity local scaling (CSLS) \cite{csls} have been proposed to mitigate the issue. 
ISF requires cross-validation of a parameter; thus we opted to use CSLS for its simpler nature. 

\subsubsection{Cross-Domain Similarity Local Scaling}

In our experiments, we use the cross-domain similarity local scaling (CSLS) metric as the ranking metric. The CSLS distance between two vectors is defined as: 
\[\textrm{CSLS}(x, y) = 2 \cos(x, y) - r_k(x) - r_k(y)  \]
where $r_k(\cdot)$ is the average cosine distance from the vector to its $k$ nearest neighbors. 
We determined that this is a better metric for ranking than simply using  cosine distance by itself. 
To save runtime, $r_k(\cdot)$ is precomputed for each candidate vector. 

\subsubsection{Edit-Distance}

We also considered using  edit distance as the ranking metric. 
Specifically, the Levenshtein distance between two words $s$ and $s'$ is  the number of single-character edits 
needed to transform $s$ into $s'$.

Although edit distance is an attractive metric due to its simplicity, it does not scale well for a large number of candidates, as its complexity is quadratic in the length of the two strings, $s$ and $s'$. 
This makes edit distance impractical for our dataset. 
Although various methods have been proposed for fast approximation of  edit distance \cite{edit1,edit2}, none have been widely adopted, nor openly implemented. 
In our experiments, we used the well-known dynamic programming algorithm \cite{wagner}. 

\section{Experiments}

We tested the accuracy and speed of our system in an information-retrieval setting on a subset of the Wiener Library's ``500 Document Collection'', which constitutes the main component of the original materials and comprises over 75,000 images that have been scanned from microfilm and microfiche copies of the originals.
We manually annotated 18 pages, totalling 4284 words. Furthermore, we created 20 random splits, each split containing 9 pages for learning the common subspace using CCA, and 9 pages for testing.

We used a set of 96 Latin characters for the PHOC representations, thus resulting in a 1440 size binary vector for each word. When using CSLS, we set ${k} = 20$.


All timing statistics are for a standard personal computer with no GPU. 

\section{Results}

We tested different versions of our system against edit distance. Table \ref{search_results} shows the performance in the information-retrieval setting. The results of the paired $t$-tests can be seen in Table \ref{t-testing}. Finally, the last column of Table \ref{fig:timing} shows the timing comparisons between the methods. The best version of our system (CCA and CSLS) is 9 times faster than edit distance, and achieves almost the same mean average precision (mAP).

The difference in performance between edit distance and our method is statistically insignificant, although edit distance performs slightly better on average. 
The differences between all other methods are statistically significant. This means that the two additions of (1) learning a common subspace to perform soft correction and (2) using the CSLS metric were effective in improving  overall performance.  

Figure~\ref{fig:qualitative} shows qualitative search results that were obtained by searching through the entire dataset. Our system is capable of finding very long words (``Nationalsozialistische''), as well as approximate matches to our query that have slightly different readings (``Rossenstrasse'' vs.\@ ``Rossenstra{\ss}e''). 
The OCR performs well on a variety of fonts. 
 
\begin{table}[t]
\caption{Search results, with and without PHOC encoding or CCA projection, and using different metrics. 
Last column gives average time on a CPU to search through the labeled test set.}
\label{search_results}
\label{fig:timing}
\begin{center}
\begin{tabular}{l|ccc|c|c}
\hline
Method & PHOC (C) & CCA (D)  & Distance (E) & mAP (\%$\pm$ s.d.) & Time (s)\\
\hline
Edit distance & \xmark & \xmark & Edit distance & $\!\!\mathbf{81.94} \pm 4.56$ & 90\\
CCA and CSLS & \cmark & \cmark & CSLS & $81.85 \pm 4.70$ & 11\\
CSLS & \cmark & \xmark & CSLS & $81.66 \pm 4.73$ & 10\\
CCA and Cosine & \cmark & \cmark & Cosine & $81.36 \pm 4.79$ & 11\\
Cosine & \cmark & \xmark & Cosine & $81.09 \pm 4.76$ & ~\textbf{8}\\
\hline
\end{tabular}
\end{center}
\end{table}

\begin{table}[t]
\caption{Statistical significance test results.}\label{t-testing}
\begin{center}
\begin{tabular}{lcc}
\hline
Pair & $p$-value & Reject $H_0$? \\
\hline
Edit distance $\leftrightarrow$ CCA and CSLS & $0.32\phantom{\;\times 10^{-4}}$ & No\\
CCA and CSLS $\leftrightarrow$ CSLS & $1.44 \times 10^{-4}$ & Yes\\
CSLS $\leftrightarrow$ CCA and Cosine & $2.06 \times 10^{-6}$ & Yes \\
CCA and Cosine $\leftrightarrow$ Cosine & $2.15 \times 10^{-6}$ & Yes\\
\hline
\end{tabular}
\end{center}
\end{table}

\section{Discussion}

\subsection{General Methods}

The methods discussed in this paper are general, and straightforward. The effects of the noisy OCR were manageable in our case. The effects of noise on text classification was discussed in \cite{agarwal}, where they concluded that up to 40\% noise was not detrimental to text classification. We determined from our manually tagged dataset that around 60\% of OCR outputs from the pretrained Tesseract model are within edit-distance 2 of the correct transcription. Despite this level of noise, we were still able to achieve compelling results. 
 
Another positive aspect for us was that  OCR works well out of the box for a variety of different languages, as well as fonts. 
This is important for our dataset because it includes German, Polish, English, and even some Hebrew texts. Within each language, there are also a variety of fonts that must be dealt with. 

\begin{figure}[t]
  \centering
  \includegraphics[width=0.8\linewidth]{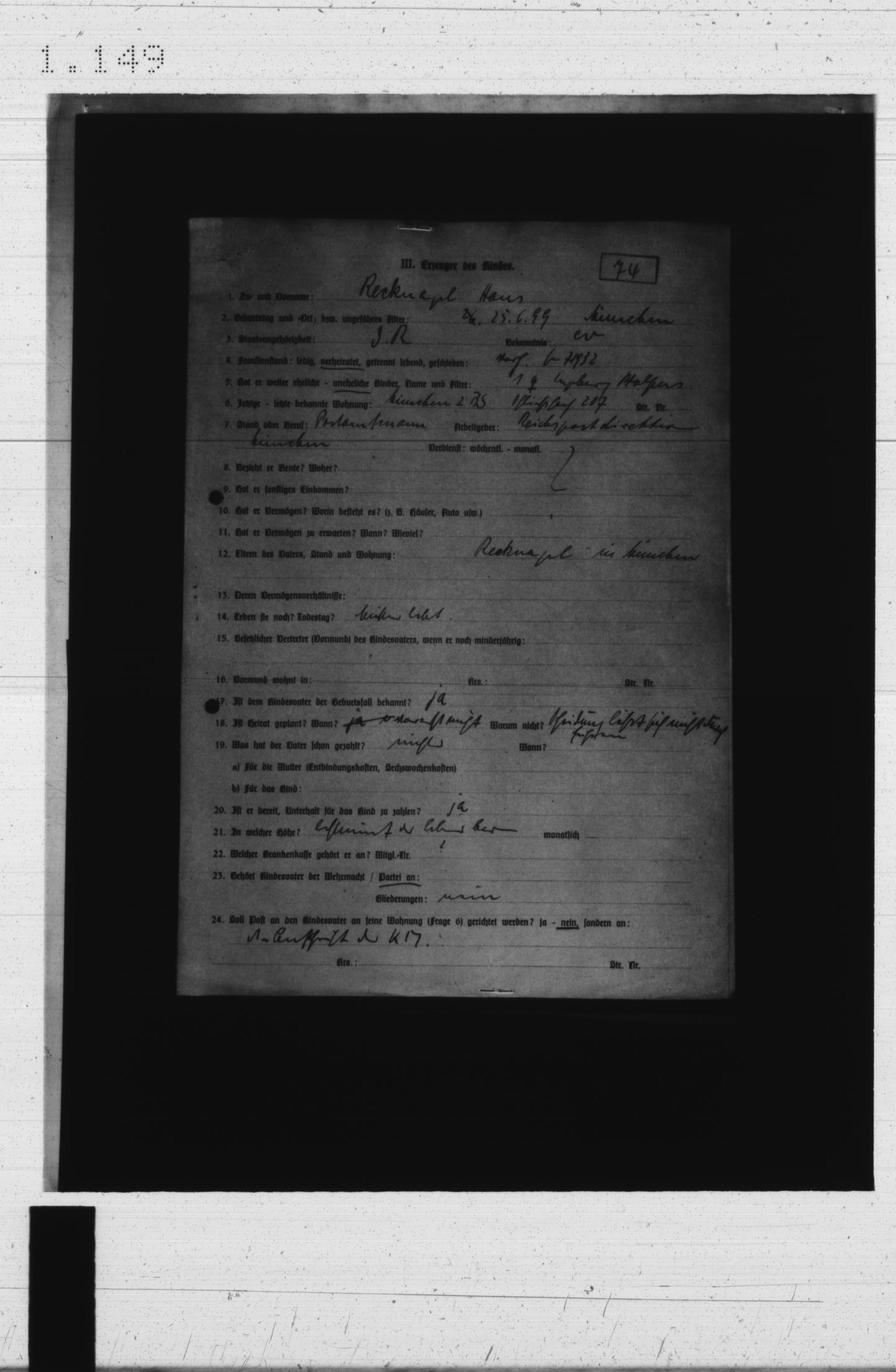}
  \caption{Example of failure case.} 
  \label{fig:dark_document}
\end{figure}

That being said, the drawback of this method is that it fails wherever the OCR fails. 
An example where  OCR fails is shown in Figure~\ref{fig:dark_document}. 
It does not predict any words in the image due to the poor quality of the image. Also, we were unable to remove the black border around the image using our current preprocessing tool, as the document is too dark. 

A careful evaluation of the quality of the OCR is necessary before using any of the proposed methods. 

\subsection{Self-Supervised OCR}

Since the OCR engine assigns a confidence score for each word that it outputs, we tried fine-tuning the OCR by using the transcriptions of its most confident outputs as additional ground truth texts. 
We took images of words that it assigned a confidence score higher than 90 percent, but were unable to improve the performance of the OCR itself. 

\subsection{Weighted Edit-Distance}

Another variant of the edit-distance takes into account the likelihood of each single character edit. 
We tried computing a confusion matrix, and using the corresponding weighted edit-distance instead of the regular edit-distance, but were not able to outperform  ``vanilla'' (uniform cost) edit distance in our experiments.

\section{Conclusion}

We have presented a fast and accurate text-based search that is easy to implement and which requires minimal fine-tuning for any given setting. 
It appears to provide an excellent balance between speed and accuracy.
It might make sense to use edit distance to rerank the top results.

We applied the system described here to all  German-language documents in the Wiener collection. 
When embedded---as  planned---in the library's search tool, this  will provide WWII scholars a valuable tool to search effectively through these important historic documents. 

State-of-the-art OCR works well out of the box for a variety of different languages and fonts. 
This is important for our dataset because it includes German, Polish, English, and Hebrew texts. Within each language, there are also a variety of styles that must be dealt with.
The method does depend on reasonable, though imperfect, OCR results.
We also experimented with word-spotting techniques, which could be combined with OCR when needed.
See \cite{Textual} for another example of leveraging poor OCR for finding related texts.

Our method is currently being applied to the other languages and to the much larger Yad Vashem collection.

\subsection*{Acknowledgments}

We thank 
Alexey Pechorin and Zahi Hazan for  technical help,
Hila Buzaglo,
Gila Michlovski,
Naama Scheftelowitz,
 Roni Stauber,
 and their teams at the university libraries
for providing the data and helping evaluate outcomes, and 
Leo Corry for making this all happen.

\bibliographystyle{plain}  
\bibliography{references}

\end{document}